\documentclass[prl,amsmath,amssymb,amsfonts,twocolumn,letterpaper,10pt,tightenlines,superscriptaddress,nofootinbib,nobalancelastpage]{revtex4}

\usepackage{tabulary}
\usepackage{hyperref}
\usepackage{qcircuit}
\usepackage{mathtools}
\usepackage{bm}
\usepackage{xcolor}
\usepackage{braket}
\usepackage{datetime}
\usepackage{stackrel}
\usepackage{stmaryrd}
\usepackage{dsfont}

\usepackage[normalem]{ulem}

\newcommand{\logicsub}{L}
\newcommand{\logic}[1]{ {#1}_\logicsub}

\newcommand{\prlsection}[1]{\vspace{0.5ex}\textbf{#1}}
\newcommand{\GKP}{\mathrm{GKP}}
\newcommand{\EC}{\mathrm{EC}}
\newcommand{\logPauli}{\op{\sigma}_\logicsub^\mu}
\DeclareMathOperator{\pdf}{pdf}

\let\arrow\vec

\allowdisplaybreaks[2]

\DeclareMathOperator{\Tr}{Tr}

\renewcommand{\Re}{\operatorname{Re}}
\renewcommand{\Im}{\operatorname{Im}}

\newcommand{\diffd}{\mathrm{d}}

\makeatletter

\newcommand{\ringplus}{\mathbin{\text{\@ringplus}}}

\newcommand{\@ringplus}{%
  \ooalign{\hidewidth\raise1.3ex\hbox{\tiny$\circ$}\hidewidth\cr$\m@th+$\cr}%
}

\newcommand{\ringminus}{\mathbin{\text{\@ringminus}}}

\newcommand{\@ringminus}{%
  \ooalign{\hidewidth\raise0.9ex\hbox{\tiny$\circ$}\hidewidth\cr$\m@th-$\cr}%
}
\makeatother

\newcommand{\tp}[0]{\mathrm{T}}

\DeclareFontFamily{U}{wncy}{}
\DeclareFontShape{U}{wncy}{m}{n}{<->wncyr10}{}
\DeclareSymbolFont{mcy}{U}{wncy}{m}{n}
\DeclareMathSymbol{\Sh}{\mathord}{mcy}{"58}

\newcommand{\negspace}{\!}
\newcommand{\lsub}[2]{{\protect\vphantom{#1}}_{#2} \negspace {#1}}
\newcommand{\rsub}[2]{{#1} \negspace {\protect\vphantom{#1}}_{#2}}
\newcommand{\lrsub}[3]{{\protect\vphantom{#1}}_{#2} \negspace {#1} \negspace {\protect\vphantom{#1}}_{#3}}

\newcommand{\ketsub}[2]{\rsub {\ket{#1}} {#2}}
\newcommand{\brasub}[2]{\lsub {\bra{#1}} {#2}}

\newcommand{\pbra}[1]{\brasub{#1} p}

\newcommand{\inprod}[2]{\left\langle {#1} | {#2} \right\rangle}
\newcommand{\inprodsubsub}[4]{\lrsub {\inprod{#1}{#2}} {#3} {#4}}

\newcommand{\outprod}[2]{\ket {#1}\!\bra {#2}}

\newcommand{\outprodsubsub}[4]{\ketsub {#1}{#3} \brasub{#2}{#4}}

\newcommand{\qoutprod}[2]{\outprodsubsub{#1}{#2}q q}

\newcommand{\reals}[0]{\mathbb{R}}

\newcommand{\integers}[0]{\mathbb{Z}}

\newcommand{\op}[1]{\hat{#1}}

\newcommand{\id}[0]{I}

\newcommand{\mat}[1]{\bm{\mathrm{#1}}}

\newcommand{\barvec}[1]{\bar{\vec{#1}}}

\renewcommand{\vec}[1]{\bm{\mathrm{#1}}}

\newcommand{\controlled}[1]{\op{\mathrm{C}}_{#1}}
\newcommand{\CZ}[0]{\controlled Z}

\usepackage{graphicx}
\graphicspath{{./figures/}}

\begin{document}

\title{All-Gaussian universality and fault tolerance with the Gottesman-Kitaev-Preskill code}

\author{ Ben Q. Baragiola}
\affiliation{Centre for Quantum Computation and Communication Technology, School of Science, RMIT University, Melbourne, Victoria, Australia}
\author{ Giacomo Pantaleoni}
\affiliation{Centre for Quantum Computation and Communication Technology, School of Science, RMIT University, Melbourne, Victoria, Australia}
\author{ Rafael N. Alexander}
\affiliation{Center for Quantum Information and Control, Department of Physics and Astronomy, University of New Mexico, Albuquerque, USA}
\author{ Angela Karanjai}
\affiliation{Centre for Engineered Quantum Systems, School of Physics, The University of Sydney, Sydney, Australia}
\author{ Nicolas C. Menicucci}
\affiliation{Centre for Quantum Computation and Communication Technology, School of Science, RMIT University, Melbourne, Victoria, Australia}

\begin{abstract}
The Gottesman-Kitaev-Preskill (GKP) encoding of a qubit within an oscillator is particularly 
appealing for fault-tolerant quantum computing  with bosons  because Gaussian operations on encoded Pauli eigenstates enable Clifford quantum computing with error correction. We show that applying GKP error correction to Gaussian input states, such as vacuum, produces distillable magic states, 
achieving universality without additional non-Gaussian elements%
. Fault tolerance is possible with sufficient squeezing and low enough external noise.  Thus, Gaussian operations are sufficient for fault-tolerant, universal quantum computing given a supply of GKP-encoded Pauli eigenstates.

\end{abstract}

\date{\today}

\maketitle

\prlsection{Introduction.}---%
%
The promise of a quantum computer lies in its ability to dramatically outpace classical computers for certain tasks~\cite{Nielsen2000}. The celebrated \emph{Threshold Theorem}~\cite{Aharonov:1997kc} proves that this feature survives even in the presence of (low enough) noise---a property called \emph{fault tolerance}, which is based on quantum error correction~\cite{Gottesman:2009ug}.

Computation using operations restricted to Pauli-eigenstate preparation, Clifford transformations, and Pauli measurements---henceforth referred to as \emph{Clifford quantum computing~(QC)}---provides all the necessary tools for quantum error correction but cannot outperform classical computation since it is efficiently simulable on a classical computer~\cite{Gottesman:1998aa}.
Universal quantum computation requires supplementing Clifford QC by a non-Clifford \emph{resource}---that is, a preparation, gate or measurement that is not an element of Clifford QC. Often this resource is a non-Pauli eigenstate, referred to as a \emph{magic state}. The union of Clifford QC and a supply of magic states is universal for quantum computing and can be made fault tolerant if the physical noise level is low enough%
 ~\cite{Bravyi:2005dx}. 

The continuous-variable~(CV) analog of Clifford QC is \emph{Gaussian QC}, which includes Gaussian state preparation, Gaussian operations (i.e., Hamiltonians quadratic in~$\op a, \op a^\dag$), and homodyne detection. CV systems arise naturally in many quantum architectures, including optical modes~\cite{chen2014experimental,YokoUkaiArms13,Yoshikawa2016,Roslund:2013cb}, microwave-cavity modes~\cite{Rosenblum2018}, and vibrational modes of trapped ions~\cite{Fluhmann2018}.  Gaussian QC lends itself to optics because the nonlinearities required are limited and low order and because homodyne detection is very high efficiency. Gaussian QC is also efficiently simulable by a classical computer~\cite{Bartlett2002} and therefore requires a non-Gaussian resource (preparation, gate, or measurement) to elevate it to universal QC~\cite{Lloyd1999}.

Fault tolerance requires discrete quantum information. \emph{Bosonic quantum-error-correcting codes} (\emph{bosonic codes} for short) embed discrete quantum information into CV systems in a way that maps general CV noise into effective noise acting on the encoded  qubits~\cite{CochMilbMunr99,Gottesman2001,Albert2018,GrimCombBara19}. 
Such codes are promising for fault-tolerant computation~\cite{Menicucci2014, Fukui2017, Vuillot2018, RoseReinMirr18} due to the built-in redundancy afforded by their infinite-dimensional Hilbert space. 
High precision controllability of optical-cavity~\cite{Ofek2016, Hu2018} and vibrational~\cite{Fluhmann2018} modes further enhances their appeal.

Using a bosonic code, one may define \emph{logical-Clifford QC}, 
comprising encoded Pauli eigenstates and logical-Clifford operations, which allows error correction  at the encoded-qubit level. This, too, is an efficiently simulable subtheory and thus requires 
a  logical-non-Clifford  element for universality. 

\emph{Our main result is  that a magic state for logical-Clifford QC (using a particular bosonic code) can be found within Gaussian QC.} Thus, the union of these two simulable subtheories is universal and---with low enough physical noise---fault tolerant.

The bosonic code that enables this is the Gottesman-Kitaev-Preskill~(GKP) encoding of a qubit into an oscillator~\cite{Gottesman2001}. This is
the only bosonic code that 
allows logical-Clifford QC and CV-level error correction to be implemented using Gaussian QC along with a supply of logical-Pauli eigenstates, which are non-Gaussian.\footnote{Note: These states cannot be prepared~\cite{Karanjai2018} using the Gaussian measurements that correspond to logical-Pauli measurements because these measurements are  destructive.} Until now, GKP logical-Clifford QC has been elevated to universal quantum computation through non-Gaussian gates (cubic phase gate) or the preparation of a logical magic state (logical Hadamard eigenstate or cubic phase state)~\cite{Gottesman2001}, all three of which require additional non-Gaussian machinery beyond the logical-Pauli eigenstate.

In what follows, we show how to produce a distillable~\cite{Bravyi:2005dx} GKP magic state using GKP error correction on a thermal state (vacuum or finite temperature), along with a complete analysis of the success probability of preparing a high-quality magic state from any given thermal state. Our result applies to both square- and hexagonal-lattice GKP  encodings~\cite{Albert2018}.
%

\prlsection{Notation and conventions}---%
%
 Here we define notation and conventions to be used throughout this Letter.
We define position~${\op q \coloneqq \tfrac {1} {\sqrt 2} (\op a + \op a^\dag)}$ and momentum~${\op p \coloneqq \tfrac {-i} {\sqrt 2} (\op a - \op a^\dag)}$ for any mode~$\op a$. This means ${[\op q, \op p] = i}$,  with a vacuum variance of~$\tfrac 1 2$ in each quadrature and ${\hbar = 1}$.

The Weyl-Heisenberg displacement operators ${\op X(s) \coloneqq e^{-is\op p}}$ and ${\op Z(s) \coloneqq e^{is\op q}}$ displace a state by~$+s$ in position and momentum, respectively. For brevity, we also define a joint displacement~${\op V(\vec s) \coloneqq \op Z(s_p) \op X(s_q)}$, where ${\vec s = (s_q, s_p)^\tp}$.

The functions ${\psi(s) \coloneqq \inprodsubsub s \psi q {}}$ and ${\tilde\psi(s) \coloneqq \inprodsubsub s \psi p {}}$ denote position- and momentum-space wave functions for a state~$\ket \psi$, respectively (tilde indicates momentum space). Any function, including wave functions, can be evaluated with respect to position, ${\varphi(\op q) \coloneqq \int \diffd s\, \varphi(s) \qoutprod s s}$, to produce an operator diagonal in the position basis---and similarly for momentum. Finally, we define~$\Sh_\Delta(x) \coloneqq \sum_{ n \in \integers } \delta(x - n\Delta)$ as a Dirac comb with spacing~$\Delta$. 

%

\prlsection{The GKP encoding}---%
%
In the original square-lattice GKP encoding~\cite{Gottesman2001}, the wave functions for the logical basis states~$\{ \ket {\logic 0}, \ket {\logic 1} \}$ are Dirac combs in position space with state-dependent offset:
%
$\psi_{j,\logicsub}(s) = \Sh_{2\sqrt \pi}(s - j \sqrt \pi)$   for ${j \in \{0,1\}}$.  
Their momentum-space wave functions are
also Dirac combs but with no offset, different spacing, and a relative phase between the spikes: 
$\tilde \psi_{j, \logicsub}(s) = \tfrac {1} {\sqrt 2} (-1)^{js/\sqrt\pi} \Sh_{\sqrt \pi}(s)$.
Note that the momentum-space spikes for $\ket {\logic 1}$ alternate sign, and those for~$\ket {\logic 0}$ are uniform.

GKP logical operators $\logic{\op{X}}$ and $\logic{\op{Z}}$ are implemented by displacements $\op{X}(\sqrt{\pi})$ and $\op{Z}(\sqrt{\pi})$, respectively, while displacements by integer multiples of $2\sqrt{\pi}$
in either quadrature leave the GKP logical subspace invariant. For later use, we define the four GKP-encoded  logical Paulis 
\begin{equation} \label{eq:encodedpaulis}
	\logPauli \coloneqq \sum_{jk} \sigma^\mu_{jk} \outprod {\logic j} {\logic k},
\end{equation} 
where $\sigma^\mu_{jk}$ is the $jk$'th element of Pauli matrix~$\mat \sigma^\mu$ (with $\mat \sigma^0 = \mat \id$). 
Note that $\logPauli$ have support only on the GKP logical subspace, while $\logic{\op{X}}$ and $\logic{\op{Z}}$ have full support.
Finally, we denote the (rank-two) projector onto the GKP logical subspace~\cite{Gottesman2001,Terhal2016}
\begin{align}
	\op \Pi_\GKP \coloneqq \op \sigma^0_\logicsub = \tilde \psi_{0,\logicsub} (\op q) \tilde \psi_{0,\logicsub} (\op p) =  \tilde \psi_{0,\logicsub} (\op p)  \tilde \psi_{0,\logicsub} (\op q)
	.
\end{align}
%

%
%

\prlsection{Kraus operator for GKP error correction.}---%
%
%
In its original formulation~\cite{Gottesman2001}, \emph{GKP error correction} is a quantum operation that corrects an encoded qubit that has acquired some noise (leakage of its wave function outside of the logical subspace) by projecting it back into the GKP logical subspace, possibly at the expense of an unintended logical operation. 
 A standard implementation of error correction strives to avoid these unintended logical operations (residual errors).  In what follows, we apply the machinery of error correction to a known Gaussian state, which means the outcome-dependent final state is known perfectly.%

GKP error correction~\cite{Gottesman2001,Menicucci2014} proceeds in two steps: First, the one quadrature is corrected, then the conjugate quadrature. 
We define the Kraus operator that corrects just the $q$ quadrature~$\op K_\EC^q(t)$  via the circuit (read right to left):
\begin{align}
\nonumber
\Qcircuit @C=0.8em @R=0.5em
{
	\qw[2] &	& \gate{K_\EC^q(t)}[2] & &
	\ustick{\text{\small in~~}}
	\\
	&		& \dstick{t} \cwx
}
\;\;&=\;\;
\Qcircuit @C=0.8em @R=0.5em
{
	\qw[2]	& & \gate{X(-t)}[1]	& \qw[1]			& \ctrl{1}[3]	& &  & 
	\ustick{\text{\small in~~}}
	\\
	&  \lstick{t}  \cw[1]		& \cctrlo{-1}[1]	 		& \push{\;\pbra{t}\;} \qw[1]	& \control \qw[1] & \push{\;\ket{\logic 0}}
	\gategroup 1 3 2 6 {.7em} {-}
}\label{eq:ECfig}
\end{align}
where the controlled operation is $\CZ = e^{i \op q \otimes \op q}$, and~$t \in \reals$ is the  measurement outcome. This circuit differs from the original~\cite{Gottesman2001} in that the correction here is a negative displacement by $t$ rather than by $t$ rounded to the nearest integer multiple of~$\sqrt \pi$. The outputs may differ by a logical operation~$\op X(\pm \sqrt \pi)$, but this is unimportant because the input state is known.

Direct evaluation shows $\op K_\EC^q(t) = \tilde \psi_{0,\logicsub}(\op{q}) \op X(-t)$. 
A similar calculation shows that the Kraus operator for correcting the~$p$ quadrature is~$\op K_\EC^p(t) = \tilde \psi_{0,\logicsub}(\op{p}) \op Z(-t)$.  
 Applying both corrections (in either order since they commute up to a phase) performs full GKP error correction: 
\begin{align}
	\op K_\EC(\vec t) = \op K_\EC^p(t_p) \op K_\EC^q(t_q) = \op \Pi_\GKP \op V(-\vec t)
	,
\end{align} 
with measurement outcomes $\vec{t} = (t_q, t_p)^\tp$. 
This Kraus operator (i)~displaces the state by an outcome-dependent amount, $\op V(-\vec t)$, and then (ii)~projects it back into the GKP logical subspace with~$\op \Pi_\GKP$.

Applying $\op K_\EC(\vec t)$ to an input state~$\op \rho_\text{in}$ produces the unnormalized state $\op {\bar \rho}(\vec t) = \op{K}_\EC(\vec t)\op{\rho}_{\rm in}\op{K}^\dagger_\EC(\vec t)%
$, where the  bar indicates  lack of normalization. The joint probability density function (pdf) for the outcomes,~$\pdf(\vec t) = \Tr [\op {\bar \rho} (\vec t)]$,  normalizes the  output  state:  $\op \rho(\vec t) = \op {\bar \rho}(\vec t) / \pdf(\vec t)$.

\prlsection{Bloch vector for the error-corrected state.}---%
%
Using the logical basis in Eqs. (\ref{eq:encodedpaulis}) we represent the output state $\op{\rho}(\vec t) = \tfrac 1 2 \sum_\mu r_\mu(\vec t) \logPauli$ by a 4-component Bloch vector $\vec{r}(\vec t)$ with outcome-dependent coefficients $r_\mu(\vec{t}) \coloneqq \Tr [\op{\rho}(\vec t) \logPauli ]$.  
 For the unnormalized state, $\bar r_0(\vec{t}) = \pdf(\vec{t})$, and for the normalized state, $r_0(\vec{t}) = 1$.
In what follows, we use the notation ${\vec r = (r_0, \arrow r)}$, where $\arrow r$ is the ordinary (3-component) Bloch vector within~$\vec r$.

\begin{figure}[t]
\centering
\includegraphics[width=\columnwidth]{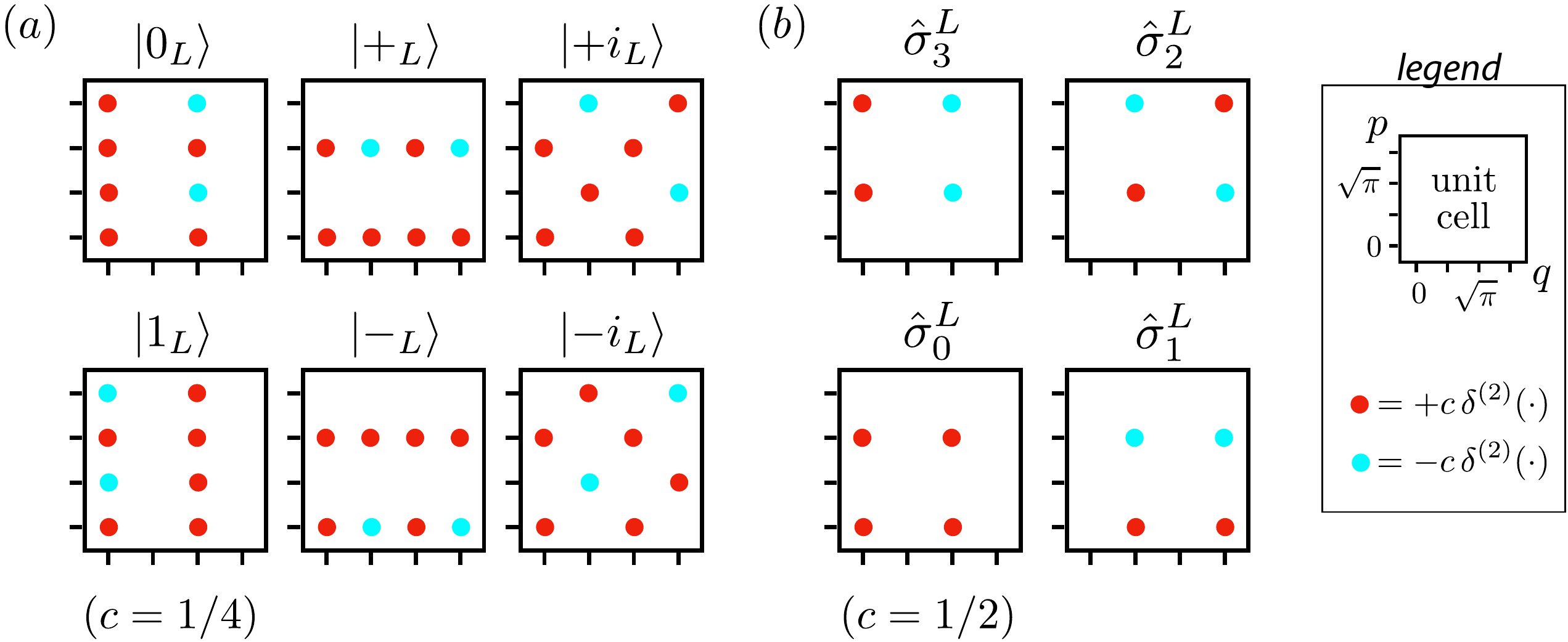} 
\caption{Wigner-function representations of the square-lattice GKP (a) Pauli eigenstates and (b) logical Pauli operators in a single unit cell of phase space with dimensions $(2\sqrt{\pi})\times(2\sqrt{\pi})$. The states are normalized to~1 over one unit cell, which determines the coefficients~$c$.%
}
\label{fig:GKPunitcell}
\end{figure} 

We employ  the Wigner functions for the logical basis states~\cite{Gottesman2001}, shown in Fig.~\ref{fig:GKPunitcell}(a), to find the Wigner functions for the  GKP-encoded Pauli operators and the GKP logical identity, Eq. (\ref{eq:encodedpaulis}). Their explicit form is 
\begin{align}
\label{eq:PauliWigner} 
	W_{\sigma^\mu_\logicsub} (\vec x)
&=
	\sum_{\vec n \in \integers^2}
	\frac {(-1)^{\vec n \cdot \barvec \ell_{\mu}}} {2}
	\delta^{(2)}
	\biggl[
		\vec x - \left( \vec n + \frac {\vec {\ell}_\mu} {2} \right) \sqrt{\pi}
	\biggr]
	, 
\end{align} 
where $\vec{x} = (q,p)^\tp$, $\vec{\ell}_0 = (0,0)^\tp$, $\vec{\ell}_1 =(1,0)^\tp$, $\vec{\ell}_2 = (1,1)^\tp$, $\vec{\ell}_3 = (0,1)^\tp$, and $\barvec{\ell}_\mu$ is just $\vec \ell_\mu$ with its entries swapped. The Wigner functions are shown in Fig.~\ref{fig:GKPunitcell}(b). 
\begin{figure*}[t]
\centering
\includegraphics[width=2\columnwidth]{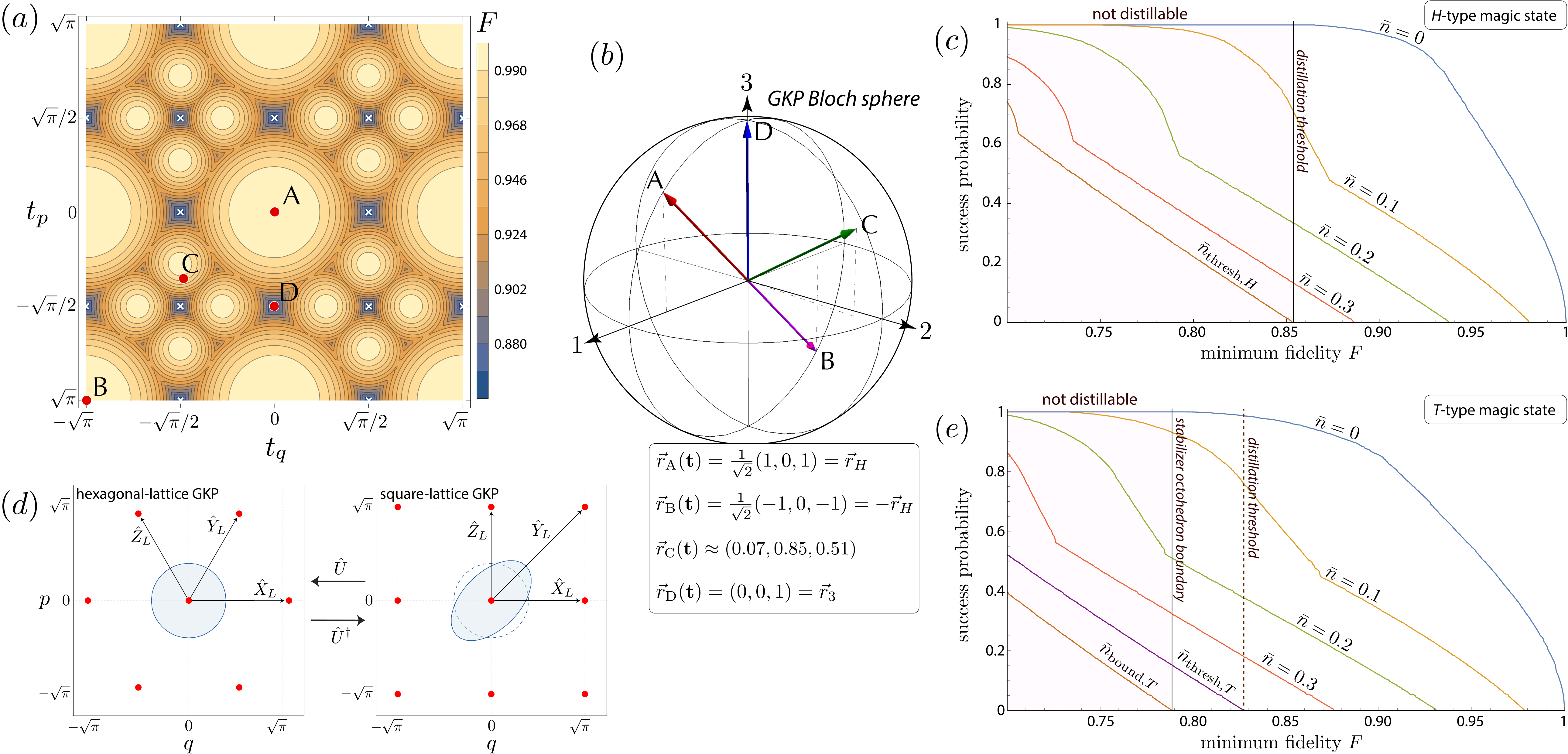} \\
\caption{(a)~GKP error correction of the vacuum: outcome-dependent fidelity~$F$ with the nearest $H$-type magic state. 
 The only outcomes that \emph{do not} yield a distillable magic state are marked with a white ``x" (these yield GKP Pauli eigenstates).   
Representative conditional Bloch vectors for outcomes (A--D) are shown on the GKP Bloch sphere in (b). 
(c)~Probability of producing an $H$-type resource state of at least fidelity~$F$ by performing GKP error correction on a thermal state of mean occupation~$\bar{n}$. Resource states with fidelity higher than the distillation threshold ($F>0.853$)~ can be distilled into higher-quality $\ket{+\logic{H}}$ states~\cite{Reichardt:2005er}. Distillation is possible for $\bar{n} < \bar{n}_{{\rm thresh},H} = 0.366$.
(d)~$\op U$ maps square-lattice GKP logical states to equivalent logical states in the hexagonal-lattice GKP encoding~\cite{Noh:2018aa}.
A vacuum state on the hexagonal lattice (1-$\sigma$ error ellipse shown in blue) is mapped to a squeezed state on the square lattice under~$\op{U}^\dagger$.
 (e)~Probability of producing a resource state distillable to $\ket{+\logic{T}^{\rm hex}}$ of at least fidelity $F$ by performing GKP$^{\rm hex}$ error correction on a thermal state of mean occupation~$\bar{n}$. Resource states whose Bloch vectors lie on or within the stabilizer octohedron ($F\leq0.789$) cannot be distilled, which occurs at $\bar{n}_{{\rm bound}, T} = 0.468$. For $T$ states, a tight distillation threshold has been proven for $F > 0.8273$~\cite{Tomas-Jochym-OConnor:2013aa}, which occurs for $\bar{n}_{{\rm thresh}, T} = 0.391$.
}
\label{fig:MagicStates}
\end{figure*} 

Since $\op \Pi_\GKP  \logPauli \op \Pi_\GKP =  \logPauli$, we skip the projection using~$\op \Pi_\GKP$ and directly calculate the unnormalized Bloch-vector components from the overlap of the  unnormalized error-corrected state~$\op {\bar \rho}(\vec t)$  with the logical Paulis. We find the overlaps in the Wigner representation:
	\begin{align} \label{eq:Blochcomponents}
		{\bar r}_\mu(\vec{t})
		&=
		\Tr[ \op {\bar \rho}(\vec t)  \logPauli ]
		=
		\Tr[ \op V(-\vec t) \op{\rho}_{\rm in} \op V^\dag(-\vec t)
		\op{\sigma}_\mu ]
\nonumber \\
		&
		=
		2 \pi \iint  d^2 \vec{x} \, W_{\rm in}(\vec{x} + \vec{t}) W_{\sigma^\mu_\logicsub}(\vec{x}),
	\end{align}
 where $W_{\rm in}(\vec{x})$ is the Wigner function of the input state $\op \rho_\text{in}$.
Note that ${\bar r}_0(\vec{t}) = \Tr [\op {\bar \rho}(\vec t)] = \pdf (\vec{t})$, which is normalized over a unit cell of size $(2 \sqrt{\pi}) \times (2 \sqrt{\pi})$ (since the full pdf is periodic).
The normalized Bloch 4-vector is  $\vec r(\vec t) \coloneqq \barvec r(\vec t) / {\bar r}_0(\vec t)$.

\prlsection{GKP error correction of Gaussian states.}---%
%
In what follows, we apply GKP error correction to a general Gaussian state---i.e., an input state whose Wigner function is $W_{\rm in}(\vec x) = G_{\vec x_0,\mat\Sigma}(\vec x)$, where $G_{\vec x_0,\mat\Sigma}$ is a normalized Gaussian with mean vector~$\vec x_0$ and covariance matrix~$\mat \Sigma$. 

Equation \eqref{eq:Blochcomponents} can be evaluated analytically when the input state is Gaussian:
	\begin{align}
	\label{eq:Blochvecexplicit}
		{\bar r}_\mu(\vec{t}) %
		 &=
		 \frac{1}{4\pi}
		 \left[G_{\vec 0 , (4 \pi \mat \Sigma)^{-1}} ( \vec{v} )\right]^{-1}
		 \Theta
		 \left(
		 \vec{v} + \frac{\barvec{\ell}_\mu}{2} , \mat{\tau} 
		 \right) \, ,
	\end{align}	
where ${\mat{\tau} = \tfrac i 2  \mat{\Sigma}^{-1}}$, ${\vec{v} = \mat{\tau} \bigl[\tfrac 1 2 \vec \ell_\mu - \tfrac 1 {\sqrt\pi} (\vec{x}_0 + \vec{t}) \bigr]}$, 
and the
Riemann (a.k.a.\ Siegel)
theta function is defined as $\Theta(\vec{z},\mat{\tau}) \coloneqq \sum_{\vec{m} \in \integers^n} \exp \left[2\pi i \left( \frac{1}{2} \vec{m}^\tp \mat{\tau} \vec{m} + \vec{m}^\tp \vec{z} \right) \right]$ 
for $\mat{\tau} \in \mathbb{H}_n$. The set $\mathbb H_n$ denotes the Siegel upper half space---i.e., the set of all complex, symmetric, $n \times n$ matrices with positive definite imaginary part (see Ref.~\cite{Deconinck:2003aa}, for example). The overall coefficient~$\tfrac {1} {4\pi}$ ensures that $\pdf(\vec t)$ 
is normalized
 over a single unit cell.

\prlsection{GKP magic states from error correction.}---%
%
GKP error correction of a Gaussian state yields a known, random state encoded in the GKP logical subspace.  Unless that state is highly mixed or too close to a logical Pauli eigenstate, it can be used as a (noisy) magic state to elevate GKP Clifford QC to fault-tolerant universal QC~\cite{Bravyi:2005dx}.  Reference~\cite{Terhal2016} suggested coupling a vacuum mode to an external qubit to perform GKP error correction and then postselecting an outcome close to $\vec t \approx \vec 0$ 
 to produce a logical $H$-type state~\cite{Bravyi:2005dx}.  In fact, neither postselection nor interaction with a material qubit is required.

With access to a supply of $\ket {\logic 0}$ states,
there is no need for any other resources except Gaussian~QC, and 
 nearly any outcome~$\vec t$ from applying GKP error correction to  the vacuum state produces a distillable $H$-type magic state~\cite{Bravyi:2005dx,Reichardt:2005er}, as shown in Fig.~\ref{fig:MagicStates}(a). This is because there are 12 $H$-type magic states (all related by Cliffords to~$\ket {+\logic H}$), and any of them will do the job~\cite{Bravyi:2005dx}. The relevant quantity is the fidelity~$F$ to the closest $H$-type state~\cite{Reichardt:2005er}. Without loss of generality, assume this is $\ket {+\logic H}$, whose Bloch 3-vector is $\arrow{r}_H = \frac{1}{\sqrt{2}} (1, 0, 1)$. (If not, apply GKP Cliffords until it is.) Then, $F = \bra {+\logic H} \op \rho(\vec t) \ket {+\logic H} = \frac{1}{2} [1 + \arrow r_H \cdot \arrow{r}(\vec t) ]$. %

Purity is not required either. Applying GKP error correction to a thermal state also succeeds with nonzero probability as long as its mean occupation number~$\bar n < 0.366 \eqqcolon \bar{n}_{{\rm thresh}, H}$; see Fig.~\ref{fig:MagicStates}(c). (A thermal state is Gaussian with $\vec{x}_0 = \vec{0}$ and $\mat{\Sigma} = (\bar{n} + \tfrac 1 2) \mat\id$ , which we plug into  Eq.~\eqref{eq:Blochvecexplicit} to produce this plot.) Since thermal states are biased towards magic states in the $xz$-plane of the Bloch sphere (see Fig.~\ref{fig:MagicStates}(d)), maximum fidelity with those magic states in the $xy$- and $yz$-plane drops below the distillation threshold first as $\bar n$ increases, leading to the kinks in Fig.~\ref{fig:MagicStates}(c).

Most high-purity, Gaussian states can be GKP-error corrected into a distillable magic state because most states do not preferentially error correct to a Pauli eigenstate.  For the vacuum, $\pdf(\vec t)$ is always between $0.066$ and~$0.094$---i.e.,  all outcomes, and thus a wide variety of states, are roughly equally likely.

%

\prlsection{ Hexagonal-lattice GKP code.}---%
%
Our  results can be extended to the hexagonal-lattice GKP code~\cite{Noh:2018aa} by simply modifying the Gaussian state to be error corrected as follows. Define $\op{U}$ as the Gaussian unitary such that $\op{U} \ket{\logic{\psi}^{\rm square}} = \ket{\logic{\psi}^{\rm hex}}$, where the logical state is the same although the encoding differs. Let $\op \rho$ be a Gaussian state to be GKP error corrected using the hexagonal lattice, with ${\vec x_0 = \vec 0}$ and covariance~$\mat \Sigma$. Then, the equivalent state to be GKP error corrected using the square lattice is $ \op \rho'_{ \rm in } = \op U^\dag \op \rho_{ \rm in }  \op U$, which is Gaussian with $\vec x_0 = \vec 0$ and covariance~$\mat \Sigma' = \mat S^{-1} \mat \Sigma \mat S^{-\tp}$~\cite{Menicucci2011}, where $\mat S = (2 \sqrt 3)^{-\tfrac 1 2} \Bigl(\begin{smallmatrix} 2 & -1 \\ 0 & \sqrt{3} \end{smallmatrix}\Bigr)$. This mapping is shown for $ \op \rho_{\rm in}  = \outprod {\text{vac}} {\text{vac}}$ in Fig.~\ref{fig:MagicStates}(d).

Using this mapping, we can get results for hexagonal-lattice GKP error correction by reusing Eq.~\eqref{eq:Blochvecexplicit} with the modified state. Vacuum and thermal states are biased towards the 
$xz$-plane of the Bloch sphere in the square-lattice encoding but unbiased with respect to all three Pauli axes in the hexagonal-lattice encoding. Thus, in Fig.~\ref{fig:MagicStates}(e), we plot the fidelity of hexagonal-lattice GKP error correction of a thermal state against $T$-type magic states~\cite{Bravyi:2005dx} such as $\ket{+\logic{T}^{\rm hex}}$, which has Bloch 3-vector $\arrow{r}_T = \frac{1}{\sqrt{3}}(1,1,1)$.

\prlsection{ Imperfect $\ket {\logic 0}$ states.}---%
%
These results can be generalized straightforwardly to the case of imperfect $\ket {\logic 0}$ states represented approximately~\cite{Menicucci2014,Motes:2017aa} as $\op K_\beta \ket {\logic 0}$, where ${\op K_\beta \coloneqq e^{-\beta \op a^\dag \op a}}$ for some~${\beta \ll 1}$ (and ignoring normalization). The approximate GKP Paulis are~$\op K_\beta \logPauli \op K_\beta$. Note that $\Tr( \op \rho  \op K_\beta \logPauli \op K_\beta ) = \Tr( \op K_\beta \op \rho \op K_\beta \logPauli)$, so we can  account for the (Gaussian) imperfections represented by~$\op K_\beta$ by applying them to the input state instead.  (Details are left to future work.) Since the fidelity requirements for magic-state distillation are orders of magnitude less than those for fault-tolerant Clifford  QC~\cite{Bravyi:2005dx,Brooks:2013wt},  any residual noise introduced by~$\op K_\beta$ will not qualitatively change  our main result.

%

\prlsection{ Error correction as heterodyne detection.}---%
%
Finally, we note that an alternate description of what we are proposing is to perform heterodyne detection (measurement in the coherent-state basis) on half of a GKP-encoded Bell pair.  This is what GKP proposed~\cite{Gottesman2001} but with photon counting replaced with heterodyne detection, which is Gaussian.  %
To see this, note that a Bell state can be written (ignoring normalisation) as $\sum_\mu \logPauli \otimes \logPauli$. Then, a coherent-state measurement on the first mode with outcome~$\alpha$ produces $\sum_\mu \Tr(\outprod \alpha \alpha \logPauli) \logPauli$ on the second mode, which agrees with Eq.~\eqref{eq:Blochcomponents} using $\op \rho_{\rm in}$ as vacuum and $\vec t = -\sqrt 2 (\Re \alpha, \Im \alpha)^\tp$. Intuitively, this is just Knill-type error correction~\cite{Knill05}, which involves teleporting the state to be corrected through an encoded Bell pair and reinterpreting vacuum teleportation as heterodyne detection.
%

%

%

\prlsection{Discussion.}---%
%
Our main result is that GKP-Clifford QC and Gaussian QC combine---with no additional non-Gaussian resources---into fault-tolerant, universal QC.  This is because $\ket {\text{vac}}$ is a distillable magic state that elevates GKP Clifford QC to fault-tolerant universality. %
Practically, this means there is no longer any need for experimentalists to pursue creating cubic phase states~\cite{Gottesman2001} if they intend to use the GKP encoding. Just  focus on making high-quality GKP $\ket {\logic 0}$ states, and the rest is all Gaussian. 
 
Fundamentally, this shows that two efficiently simulable subtheories, when used together, are universal and fault tolerant. This is  straightforward  for qubits: just combine Clifford QC based on different Pauli frames---e.g., ${\{\op X, \op Y, \op Z\}}$ and ${\{\op H, \op Y, \op Z \op H \op Z\}}$---since stabilizer states of one are magic states for the other. But this was neither known nor appreciated for CV systems until now.

We have demonstrated the ``magic'' of error correction by deploying it in a nonstandard way to produce resource states from a known, easy-to-prepare state. 
The ``wilderness space'' outside a bosonic code's logical subspace may be rich in other resources, too---e.g., providing  the  means to produce other logical states or perform logical operations more easily than would be possible by restricting to the logical subspace. 
 This feature is likely to extend beyond GKP to other bosonic codes such as rotation-symmetric codes~\cite{GrimCombBara19}, experimentally proven cat codes~\cite{Ofek2016}, and multi-mode GKP codes~\cite{Albert2018, Noh:2018aa}.   

%

%
%

\prlsection{Acknowledgments.}---%
%
We thank Andrew Landahl for discussions.
R.N.A.\ is supported by National Science Foundation Grant No.\ PHY-1630114. A.K.\ is supported by the Australian Research Council Centre of Excellence for Engineered Quantum Systems (Project No. CE170100009).
This work is supported by the Australian Research Council Centre of Excellence for Quantum Computation and Communication Technology (Project No.\ CE170100012).

\bibliographystyle{bibstyleNCM_papers_notitle}
\bibliography{MenicucciPapersRefs,bibliography}

\clearpage

\end{document}